\documentclass[iop,apj,twocolumn]{emulateapj}
\usepackage{apjfonts}
\newcommand{\jopp}{J. Plasma Phys.}
\newcommand{\rslpsa}{Proc. R. Soc. A}
\newcommand{\lrsp}{Living Rev. Sol. Phys.}
\newcommand{\pof}{Phys. Fluids}
\newcommand{\pop}{Phys. Plasmas}

\newcommand{\jltp}{J. Low Temp. Phys.}
\newcommand{\ang}{Ann. Geophys.}

\newcommand{\Alfven}{Alfv\'{e}n}
\newcommand{\Alfvenic}{Alfv\'{e}nic}
\newcommand{\para}{\parallel}
\newcommand{\PA}{$P_{\perp}/P_{\para}$}
\newcommand{\kA}{$k_{\perp}/k_{\para}$}

\shorttitle{PLASMA TURBULENCE ANISOTROPY}
\shortauthors{CHEN ET AL.}
\slugcomment{Submitted 2009 September 14; accepted 2010 January 27; published 2010 February 19}
\journalinfo{The Astrophysical Journal Letters, 711:L79--L83, 2010 March 10}
\def\andname{}

\begin{document}

\begin{minipage}{2\linewidth}
\vspace{1cm}
\scriptsize \copyright\ 2010. The American Astronomical Society. All rights reserved. Printed in the U.S.A. \normalsize
\end{minipage}

\title{Interpreting Power Anisotropy Measurements in Plasma Turbulence}

\author{C.~H.~K.~Chen\altaffilmark{1}, R.~T.~Wicks\altaffilmark{1}, T.~S.~Horbury\altaffilmark{1}, and A.~A.~Schekochihin\altaffilmark{2}}
\affil{$^1$The Blackett Laboratory, Imperial College London, London SW7 2AZ, UK; christopher.chen03@imperial.ac.uk}
\affil{$^2$Rudolf Peierls Centre for Theoretical Physics, University of Oxford, Oxford OX1 3NP, UK}

\begin{abstract}
A relationship is derived between power anisotropy and wavevector anisotropy in turbulent fluctuations. This can be used to interpret plasma turbulence measurements, for example, in the solar wind. If fluctuations are spatially anisotropic, then the ion gyroscale break point in measured spectra in the directions parallel and perpendicular to the magnetic field would not occur at the same frequency, and similarly for the electron gyroscale break point. This is an important consideration when interpreting solar wind measurements in terms of anisotropic turbulence theories. Model magnetic field power spectra are presented assuming a cascade of critically balanced \Alfven\ waves in the inertial range and kinetic \Alfven\ waves in the dissipation range. The variation of power anisotropy with scale is compared to existing solar wind measurements, and the similarities and differences are discussed.
\end{abstract}


\section{Introduction}

Plasma turbulence is observed to be anisotropic with respect to the magnetic field direction. For example, in the solar wind, the observed power and scaling of turbulent fluctuations vary depending on the angle between the local mean field and the sampling direction \citep{bieber96,horbury08,podesta09a,osman09a}. Correlation functions have also been observed to be anisotropic in the solar wind \citep{crooker82,matthaeus90,osman07,weygand09} and laboratory measurements \citep{robinson71,zweben79}.

Recent theories of plasma turbulence assume anisotropy \citep{goldreich95,boldyrev06,galtier06a,lithwick07,gogoberidze07,chandran08,beresnyak08,podesta09b,schekochihin09} and anisotropic energy transfer has been seen in simulations \citep{shebalin83,cho00,maron01,cho02,cho04,cho09}. The theories usually describe the anisotropy in terms of the fluctuation wavenumbers parallel and perpendicular to the mean magnetic field direction, $k_{\para}$ and $k_{\perp}$. For example, \citet{goldreich95} used the ``critical balance'' assumption to obtain $k_{\para} \sim k_{\perp}^{2/3}$ for magnetohydrodynamic (MHD) turbulence and, more generally, theories often assume $k_{\perp} \gg k_{\para}$. In the solar wind, however, it is the anisotropy in power at a fixed scale that is often measured for practical reasons, rather than the spatial anisotropy of the fluctuations. 

In Section \ref{sec:pavska}, the relationship between power anisotropy and wavevector anisotropy is derived. A critically balanced model is then presented in Section \ref{sec:cbspectrum} to illustrate that a break point in an anisotropic spectrum may occur at different scales when the reduced spectrum is observed in different directions. The implications of these two sections for recent solar wind measurements are discussed in Section \ref{sec:measurements}.

\section{Power Anisotropy and Wavevector Anisotropy}
\label{sec:pavska}

The correlation function of a turbulent field, for example, the magnetic field, $\mathbf{B}$, can be defined as $C(\mathbf{x}) = \left\langle \mathbf{B}(\mathbf{r}+\mathbf{x}) \cdot \mathbf{B}(\mathbf{r})\right\rangle$, where the angular brackets denote an ensemble average over positions $\mathbf{r}$. The three-dimensional energy spectrum can then be defined as the Fourier transform of the correlation function, $E(\mathbf{k})=\int C(\mathbf{k})e^{-i\mathbf{k}\cdot\mathbf{x}}d^3\mathbf{x}$.

A single spacecraft in the solar wind measures the turbulent field as a function of time, $\mathbf{B}(t)$. Since the solar wind velocity, $\mathbf{v}_{\text{sw}}$, is much larger than the wave speed (often taken as the \Alfven\ speed in the inertial range), Taylor's hypothesis \citep{taylor38} is usually well satisfied, meaning that the measured time variations correspond to spatial fluctuations in the plasma, $\Delta\mathbf{x}=-\mathbf{v}_{\text{sw}}\Delta t$. Because a single spacecraft gives a one-dimensional cut through the plasma, the full three-dimensional spectrum cannot be measured but instead a reduced version is obtained \citep{fredricks76}. This reduced spectrum, defined as $P(k)=\int E(\mathbf{k'})\delta(k-\mathbf{k'}\cdot\mathbf{\hat{v}}_{\text{sw}})d^3\mathbf{k'}$, where $\mathbf{\hat{v}}_{\text{sw}}$ is the solar wind direction unit vector, is the three-dimensional spectrum integrated over the directions perpendicular to the measuring direction. Assuming axisymmetry about the magnetic field, this reduced power spectrum also depends on the angle, $\theta$, of the field to the one-dimensional measurement direction, $P(k,\theta)$. In Cartesian coordinates, this can be written as a dependence on the parallel and perpendicular wavenumbers, $P(k_{\para},k_{\perp})$.

\begin{figure}
\epsscale{1.18}
\plotone{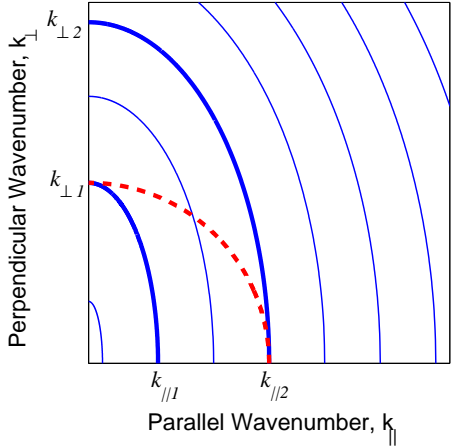}
\caption{\label{contours}Schematic of reduced power contours (solid blue lines) elongated in the field parallel direction. Power anisotropy measurements are made at a fixed scale (dashed red line).}
\end{figure}

Figure \ref{contours} is a schematic of reduced power contours with respect to the parallel and perpendicular wavenumbers. Power anisotropy is usually measured at a fixed scale, indicated by the red dashed line which is at a fixed radius from the origin. At different points along this line, a different reduced power is sampled; this effect is readily seen in the solar wind \citep{bieber96,horbury08,podesta09a,osman09a}. A relationship between power anisotropy and wavevector anisotropy will now be derived. Note that the derivation does not depend on any particular contour shape; the elliptical shapes in the figure are for illustrative purposes only.

Let us consider two contours of size $(k_{\para 1},k_{\perp 1})$ and $(k_{\para 2},k_{\perp 2})$ such that $k_{\para 2}=k_{\perp 1}$ (Figure \ref{contours}). We will assume that power anisotropy is being measured at this wavenumber, $k=k_{\para 2}=k_{\perp 1}$. Let reduced power in the parallel and perpendicular directions be defined as $P_{\para}=P(k_{\para},0)$ and $P_{\perp}=P(0,k_{\perp})$, and $\alpha$ be the scaling exponent in the perpendicular direction, $P_{\perp}\sim k_{\perp}^{-\alpha}$. Dividing the equations for $P_{\perp}$ for each contour we get
\begin{equation}
\label{kperpscaling}
\frac{P_{\perp 1}}{P_{\perp 2}} = \left( \frac{k_{\perp 2}}{k_{\perp 1}} \right) ^{\alpha}.
\end{equation}
By the definition of a contour, $P_{\perp 2}=P_{\para 2}$ and from the above definition of the two contours, $k_{\para 2}=k_{\perp 1}$. Substituting these into Equation (\ref{kperpscaling}) gives
\begin{equation}
\label{kperpscaling2}
\frac{P_{\perp 1}}{P_{\para 2}} = \left( \frac{k_{\perp 2}}{k_{\para 2}} \right) ^{\alpha}.
\end{equation}
Since $P_{\perp 1}/P_{\para 2}$ is just the power anisotropy at fixed wavenumber $k$ (as in solar wind measurements), the numeric subscripts may be dropped. Rearranging Equation (\ref{kperpscaling2}) we get
\begin{equation}
\label{pa2ka2}
\left(\frac{k_{\perp}}{k_{\para}}\right)_2 = \left( \frac{P_{\perp}}{P_{\para}} \right) ^{\frac{1}{\alpha}}.
\end{equation}
This relationship is independent of the scaling of the parallel spectrum and allows us to calculate the wavevector anisotropy of contour 2 from a measurement of \PA. A similar relationship can be derived for contour 1,
\begin{equation}
\label{pa2ka1}
\left(\frac{k_{\perp}}{k_{\para}}\right)_1 = \left( \frac{P_{\perp}}{P_{\para}} \right) ^{\frac{1}{\beta}},
\end{equation}
where $\beta$ is the scaling exponent of the parallel reduced spectrum, $P_{\para}\sim k_{\para}^{-\beta}$. Although it is possible to infer the wavevector anisotropy from the interpolation of measurements such as Figure 1 of \citet{horbury08}, the relationship given here allows it to be found from the power anisotropy, a quantity more easily measurable in the solar wind.

The $k_{\perp}$ and $k_{\para}$ of turbulence theories usually describe typical length scales associated with the fluctuations. It is usually assumed that second-order statistics, such as power, relate to these quantities, for example \citet{cho00} state that in their simulations, contours of second-order structure functions ``reflect the shapes of the eddies.'' Under this assumption, the wavevector anisotropy of power contours can be thought to describe ``typical'' wavevector anisotropy of the fluctuations.

\section{Form of the Critical Balance Reduced Power Spectrum}
\label{sec:cbspectrum}

The ``critical balance'' assumption states that in a turbulent \Alfven\ wave (AW) cascade, the linear wave timescale and the nonlinear energy transfer timescale are comparable. It was introduced explicitly by \citet{goldreich95} and anticipated in the work of \citet{higdon84}. When applied to inertial range MHD turbulence, the spectral index of the reduced spectrum in the perpendicular direction is $-5/3$ and the wavevector scaling is $k_{\para} \sim k_{\perp}^{2/3}$. A reduced spectral index of $-2$ in the parallel direction follows from these statements.

There is evidence in the solar wind inertial range for both the \Alfvenic\ nature of the turbulence \citep[e.g.][]{belcher71,horbury95,bale05} and the anisotropic scaling \citep{horbury08}. It appears that this scaling is only detectable when observing with respect to the scale-dependent \emph{local} mean magnetic field \citep[e.g.][]{cho00,horbury08,beresnyak09}, i.e., the mean field at the scale of each fluctuation being measured, rather than a global large-scale average field. Although MHD is a fluid theory, \citet{schekochihin09} have shown that reduced MHD, an anisotropic limit of MHD containing the \Alfvenic\ fluctuations, can be derived for a collisionless plasma at scales larger than the ion gyroradius. This may explain why the MHD scalings are seen in the collisionless solar wind.

At scales smaller than the ion gyroradius, commonly termed the ``dissipation range,'' there is evidence for kinetic \Alfven\ waves (KAWs) \citep{bale05,sahraoui09}. These are linear modes of electron reduced MHD, an anisotropic theory derived for collisionless plasmas at scales between the electron and ion gyroradii \citep{schekochihin09}. When the critical balance assumption is applied to a KAW cascade, the wavevector scaling becomes $k_{\para} \sim k_{\perp}^{1/3}$ and the predicted spectral indices for the magnetic field are $-7/3$ in the perpendicular direction and $-5$ in the parallel direction \citep{cho04,schekochihin09}.

In this theoretical framework, the break between the inertial range and the dissipation range is predicted to be at $k_{\perp}\rho_i \sim 1$, where $\rho_i$ is the ion gyroradius, but if the fluctuations here are anisotropic then their parallel length should be larger, $k_{\para}\rho_i < 1$. This would imply that the observed break points in solar wind measurements are at different spacecraft frequencies for the reduced spectra in the parallel and perpendicular directions. A similar effect would be expected at the electron break scale, $k_{\perp}\rho_e \sim 1$, where the difference in break frequency between the spectra in the parallel and perpendicular directions may be even greater if the anisotropy continues to increase throughout the dissipation range.

\begin{figure}
\plotone{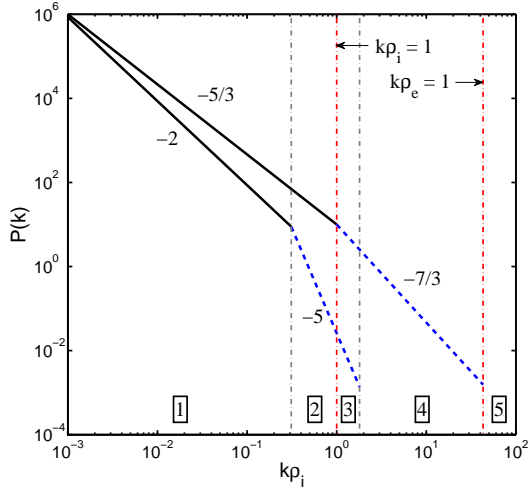}
\caption{\label{modelspectra}Schematic of magnetic field reduced power spectra in the parallel and perpendicular directions for critically balanced AW (solid black lines) and KAW (dashed blue lines) turbulence.}
\end{figure}

Since the observed break points would be at different scales for spectra in the parallel and perpendicular directions, a schematic of the reduced spectra can be divided into five ranges (Figure \ref{modelspectra}). In range 1 both spectra display AW scaling; in range 2 the spectrum in the parallel direction has KAW scaling and the spectrum in the perpendicular direction has AW scaling; in range 3 both spectra display KAW scaling; in range 4 the spectrum in the parallel direction is below the electron break scale and the spectrum in the perpendicular direction has KAW scaling; in range 5 both spectra are below the electron break scale. Predictions for fluctuations smaller than the electron scale do exist but have not been included here. Gyrokinetic theory predicts scalings for an electron-entropy cascade, valid for $k_{\perp}\rho_e \gg 1$ \citep{schekochihin09}, however it has been suggested that it is not applicable to the solar wind in this range \citep{howes08a}. In Figure \ref{modelspectra}, the (logarithm of the) power anisotropy can be thought of as the vertical distance between the spectra and the (logarithm of the) wavevector anisotropy as the horizontal distance.

\begin{figure}
\plotone{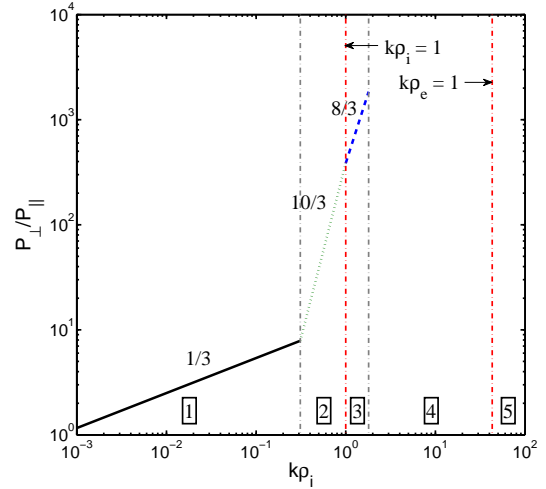}
\caption{\label{modelpa}Schematic of reduced power anisotropy as a function of scale for critically balanced AW and KAW turbulence.}
\end{figure}

The scalings for each of the ranges in Figure \ref{modelspectra} are given in Table \ref{tab:scaling}. Also listed is the scaling of \PA , which follows directly from that of $P_{\perp}$ and $P_{\para}$ and is shown in Figure \ref{modelpa}. In the inertial range \PA\ scales as $k^{1/3}$ which steepens to $k^{10/3}$ when $P_{\para}$ reaches the ion break scale and then becomes shallower at $k^{8/3}$ when $P_{\perp}$ reaches the ion break scale.

\begin{deluxetable}{cccc}
\tablecaption{\label{tab:scaling}Power Scaling Exponent Predictions for Critically Balanced AW and KAW Turbulence}
\tablehead{\colhead{Range} & \colhead{$P_{\perp}$ Scaling} & \colhead{$P_{\para}$ Scaling} & \colhead{\PA\ Scaling}}
\startdata
1 & -5/3 & -2 & 1/3 \\
2 & -5/3 & -5 & 10/3 \\
3 & -7/3 & -5 & 8/3 \\
4 & -7/3 & $\cdot\cdot\cdot$ & $\cdot\cdot\cdot$ \\
5 & $\cdot\cdot\cdot$ & $\cdot\cdot\cdot$ & $\cdot\cdot\cdot$
\enddata
\end{deluxetable}

The width of the KAW range in the $P_{\perp}$ spectrum is predicted to be $\rho_i/\rho_e=\sqrt{T_i m_i/T_e m_e}$, where $m_i$ and $m_e$ are the ion and electron masses, and for the model spectra in Figures \ref{modelspectra} and \ref{modelpa} the temperatures have been assumed equal, $T_i=T_e$. Although $T_i/T_e$ is of order unity in the solar wind, there is some variation \citep{bruno05} so the extent of the possible KAW range may vary. The width of each of the ranges in Figures \ref{modelspectra} and \ref{modelpa} also depends on the amount of anisotropy present. For example, as the anisotropy at the ion break scale increases, the size of range 2 increases but the size of range 3 decreases. It may even be the case that if the anisotropy is very strong at the ion break scale then range 3 may not be present at all, meaning it would not be possible to measure KAW scaling in $P_{\perp}$ and $P_{\para}$ at the same frequency.

One of the assumptions used when constructing these model spectra was that they contain no additional energy injection or loss. As noted in \citet{schekochihin09}, at the ion break scale some energy may be transferred from the \Alfvenic\ cascade channel to a purely electrostatic ``entropy cascade.'' The amount of energy transferred, if any, is unknown so cannot be included in our model spectra here. It is also possible that power may be injected into the cascades from other sources such as plasma instabilities, for example, the firehose and mirror instabilities, which evidence suggests may be important in the solar wind \citep{bale09}. Including effects such as these in this model is beyond the scope of this Letter. 

It should also be noted that only the \Alfvenic\ part of the cascade in the inertial range is dealt with here. This is relevant to the solar wind, which is primarily \Alfvenic\ in nature \cite[e.g.][]{belcher71,bale05,bruno05} and in which any compressive (non-\Alfvenic) fluctuations are, on theoretical grounds, not thought to interfere with the \Alfvenic\ cascade \citep{cho03,schekochihin09}.

\section{Comparison to Solar Wind Measurements}
\label{sec:measurements}

The only published measurement of the variation of power anisotropy, \PA, in the solar wind across both inertial and dissipation ranges, which we are aware of, is in the lower panel of Figure 7 of \citet{podesta09a}. The scaling arguments in Section \ref{sec:pavska} can be applied to these measurements to obtain estimates of the wavevector anisotropy at various scales. For example, at the high-frequency end of the inertial range, at $\approx$ 0.2 Hz, $P_{\perp}/P_{\para} \approx 7$ which, using $\alpha=5/3$ in Equation (\ref{pa2ka2}), means $k_{\perp}/k_{\para} \approx 3$. A value of $\alpha=5/3$ was used for this calculation since it is the prediction of critical balance MHD turbulence \citep{goldreich95} and is close to the measured value of 1.65 for this data interval \citep{podesta09a}. In general, break points in spectra seem to have a rollover rather than a clean break in scaling so this result is approximate. It is also possible that since $P_{\para}$ is measured in the bin 0 -- 6$^{\circ}$ and not at exactly $0^{\circ}$, values of \PA\ and therefore \kA\ may be underestimates. The value obtained here, $k_{\perp}/k_{\para} \approx 3$, was used to set the anisotropy at the ion break scale in Figures \ref{modelspectra} and \ref{modelpa}. 

Figure \ref{modelpa} of this Letter can be compared to Figure 7 of \citet{podesta09a} in which the frequency corresponding to $k\rho_i=1$ is $\approx$ 0.5 Hz. In the inertial range, both graphs have a shallow slope with the measurement steeper than the prediction, although the uncertainties in the measured slope may be significant as discussed in \citet{podesta09a}. Between 0.4 Hz and 1.0 Hz, the measured power anisotropy increases with a steep slope. The scales at which this happens approximately correspond to ranges 2 and 3 in Figure \ref{modelpa}. It may be the case, therefore, that the steep slope of \PA\ seen in dissipation range solar wind measurements is due to critical balance scaling, in particular the steep scaling of the KAW $P_{\para}$ spectrum.

One difference between the figures is that the decrease in power anisotropy between 0.2 Hz and 0.4 Hz in the measurement is not present in Figure \ref{modelpa}. As discussed in \citet{podesta09a}, this is caused by an increase in the parallel power and may be due to parallel waves, for example, from plasma instabilities. For frequencies above 1.0 Hz the measured anisotropy decreases, which is due to the flattening of the $P_{\para}$ spectrum. The scale at which this begins (1.0 Hz) is close to the predicted electron break scale for the $P_{\para}$ spectrum, although without knowing the electron gyroradius, the exact location of this is not clear. One possibility, therefore, is that the cause of this decrease in \PA\ may be due to the $P_{\para}$ spectrum flattening above the electron break scale. Another possibility is that the cyclotron resonance may have been reached here, causing a change in behavior \citep{howes08a}. At these high frequencies, however, measurement effects, such as magnetometer noise, may be important and one must be cautious when drawing any conclusions from this range.

Extrapolating the model spectra to larger scales, it can be seen from Figures \ref{modelspectra} and \ref{modelpa} that power (and wavevector) isotropy is reached at $k\rho_i \approx 10^{-3}$. This corresponds to a length around $10^6$ km or an observed spacecraft frequency of $5 \times 10^{-4}$ Hz. This is close to the observed break between the low-frequency $f^{-1}$ power law and the inertial range \citep[e.g.][]{bavassano82a,bruno05} and also the solar wind correlation length \citep[e.g.][]{matthaeus82a}, scales usually associated with the outer scale of the turbulence.

The solar wind spectrum has recently been observed at scales near $k\rho_e=1$. \citet{alexandrova09} suggest that just above this scale there is an exponential falloff in the spectrum, and \citet{sahraoui09} suggest that below it, there is a further steeper power law. Both of these studies involved solar wind intervals where the magnetic field was not aligned with the solar wind direction so that one would expect to see the spectrum in the perpendicular direction. \citet{sahraoui09} also plot the spectrum of the parallel component of the magnetic field, which, it should be pointed out for clarity, is not the same as the spectrum in the parallel direction.

\section{Summary and Conclusions}

In Section \ref{sec:pavska}, a relationship was derived that allows the turbulent wavevector anisotropy, \kA, to be inferred from power anisotropy measurements, \PA. This is independent of any particular turbulence theory and only assumes power-law scaling in the parallel and perpendicular directions. Using this relation and existing solar wind measurements \citep{podesta09a} the wavevector anisotropy near the ion break scale was estimated to be $k_{\perp}/k_{\para}\approx 3$, although this may be an underestimate due to the finite angular resolution of the measurements.

Model spectra of critically balanced AWs (for the inertial range) and KAWs (for the dissipation range) were presented to illustrate that break points do not occur at the same scale in the observed spectra in the parallel and perpendicular directions if the turbulence is anisotropic. The variation of \PA\ with scale was calculated from these model spectra resulting in five ranges, three of which have predictions of how \PA\ scales: a 1/3 range, a 10/3 range, and an 8/3 range. If the wavevector anisotropy is significant then some of these ranges are small and may not even be present. Some of these features can be seen in the measurements of \citet{podesta09a}. The main difference is the extra parallel power at the ion break scale seen in the measurements, which may be due to energy injection mechanisms at the ion gyroscale.

Although critically balanced AWs and KAWs were used in Section \ref{sec:cbspectrum}, the ideas also apply to anisotropic theories of plasma turbulence in general. Some of these, for example, are a nonlocal cascade model \citep{gogoberidze07}, turbulence with dynamic alignment \citep{boldyrev06,podesta09b}, wave turbulence in Hall MHD \citep{galtier06a}, and imbalanced turbulence \citep{lithwick07,beresnyak08,chandran08}. The anisotropy relationship derived in Section \ref{sec:pavska} and the observational considerations discussed in Sections \ref{sec:cbspectrum} and \ref{sec:measurements} are also applicable to these theories and any possible extensions of them into the dissipation range.

\acknowledgments
This work was funded by STFC and the Leverhulme Trust International Academic Network for Magnetized Plasma Turbulence. C.~Chen acknowledges helpful discussions with T.~Yousef and A.~Mallet and useful comments from an anonymous referee.


\begin{thebibliography}{40}
\expandafter\ifx\csname natexlab\endcsname\relax\def\natexlab#1{#1}\fi

\bibitem[{{Alexandrova} {et~al.}(2009){Alexandrova}, {Saur}, {Lacombe},
  {Mangeney}, {Mitchell}, {Schwartz}, \& {Robert}}]{alexandrova09}
{Alexandrova}, O., {Saur}, J., {Lacombe}, C., {Mangeney}, A., {Mitchell}, J.,
  {Schwartz}, S.~J., \& {Robert}, P. 2009, \prl, 103, 165003

\bibitem[{{Bale} {et~al.}(2009){Bale}, {Kasper}, {Howes}, {Quataert}, {Salem},
  \& {Sundkvist}}]{bale09}
{Bale}, S.~D., {Kasper}, J.~C., {Howes}, G.~G., {Quataert}, E., {Salem}, C., \&
  {Sundkvist}, D. 2009, \prl, 103, 211101

\bibitem[{{Bale} {et~al.}(2005){Bale}, {Kellogg}, {Mozer}, {Horbury}, \&
  {Reme}}]{bale05}
{Bale}, S.~D., {Kellogg}, P.~J., {Mozer}, F.~S., {Horbury}, T.~S., \& {Reme},
  H. 2005, \prl, 94, 215002

\bibitem[{{Bavassano} {et~al.}(1982){Bavassano}, {Dobrowolny}, {Mariani}, \&
  {Ness}}]{bavassano82a}
{Bavassano}, B., {Dobrowolny}, M., {Mariani}, F., \& {Ness}, N.~F. 1982, \jgr,
  87, 3617

\bibitem[{{Belcher} \& {Davis}(1971)}]{belcher71}
{Belcher}, J.~W., \& {{Davis},\ L.,\ Jr.} 1971, \jgr, 76, 3534

\bibitem[{{Beresnyak} \& {Lazarian}(2008)}]{beresnyak08}
{Beresnyak}, A., \& {Lazarian}, A. 2008, \apj, 682, 1070

\bibitem[{{Beresnyak} \& {Lazarian}(2009)}]{beresnyak09}
---. 2009, \apj, 702, 460

\bibitem[{{Bieber} {et~al.}(1996){Bieber}, {Wanner}, \& {Matthaeus}}]{bieber96}
{Bieber}, J.~W., {Wanner}, W., \& {Matthaeus}, W.~H. 1996, \jgr, 101, 2511

\bibitem[{{Boldyrev}(2006)}]{boldyrev06}
{Boldyrev}, S. 2006, \prl, 96, 115002

\bibitem[{{Bruno} \& {Carbone}(2005)}]{bruno05}
{Bruno}, R., \& {Carbone}, V. 2005, \lrsp, 2, 4

\bibitem[{{Chandran}(2008)}]{chandran08}
{Chandran}, B.~D.~G. 2008, \apj, 685, 646

\bibitem[{{Cho} \& {Lazarian}(2003)}]{cho03}
{Cho}, J., \& {Lazarian}, A. 2003, \mnras, 345, 325

\bibitem[{{Cho} \& {Lazarian}(2004)}]{cho04}
---. 2004, \apjl, 615, L41

\bibitem[{{Cho} \& {Lazarian}(2009)}]{cho09}
---. 2009, \apj, 701, 236

\bibitem[{{Cho} {et~al.}(2002){Cho}, {Lazarian}, \& {Vishniac}}]{cho02}
{Cho}, J., {Lazarian}, A., \& {Vishniac}, E.~T. 2002, \apj, 564, 291

\bibitem[{{Cho} \& {Vishniac}(2000)}]{cho00}
{Cho}, J., \& {Vishniac}, E.~T. 2000, \apj, 539, 273

\bibitem[{{Crooker} {et~al.}(1982){Crooker}, {Siscoe}, {Russell}, \&
  {Smith}}]{crooker82}
{Crooker}, N.~U., {Siscoe}, G.~L., {Russell}, C.~T., \& {Smith}, E.~J. 1982,
  \jgr, 87, 2224

\bibitem[{{Fredricks} \& {Coroniti}(1976)}]{fredricks76}
{Fredricks}, R.~W., \& {Coroniti}, F.~V. 1976, \jgr, 81, 5591

\bibitem[{{Galtier}(2006)}]{galtier06a}
{Galtier}, S. 2006, \jltp, 145, 59

\bibitem[{{Gogoberidze}(2007)}]{gogoberidze07}
{Gogoberidze}, G. 2007, \pop, 14, 022304

\bibitem[{{Goldreich} \& {Sridhar}(1995)}]{goldreich95}
{Goldreich}, P., \& {Sridhar}, S. 1995, \apj, 438, 763

\bibitem[{{Higdon}(1984)}]{higdon84}
{Higdon}, J.~C. 1984, \apj, 285, 109

\bibitem[{{Horbury} {et~al.}(1995){Horbury}, {Balogh}, {Forsyth}, \&
  {Smith}}]{horbury95}
{Horbury}, T.~S., {Balogh}, A., {Forsyth}, R.~J., \& {Smith}, E.~J. 1995, \grl,
  22, 3405

\bibitem[{{Horbury} {et~al.}(2008){Horbury}, {Forman}, \&
  {Oughton}}]{horbury08}
{Horbury}, T.~S., {Forman}, M., \& {Oughton}, S. 2008, \prl, 101, 175005

\bibitem[{{Howes} {et~al.}(2008){Howes}, {Cowley}, {Dorland}, {Hammett},
  {Quataert}, \& {Schekochihin}}]{howes08a}
{Howes}, G.~G., {Cowley}, S.~C., {Dorland}, W., {Hammett}, G.~W., {Quataert},
  E., \& {Schekochihin}, A.~A. 2008, \jgr, 113, 5103

\bibitem[{{Lithwick} {et~al.}(2007){Lithwick}, {Goldreich}, \&
  {Sridhar}}]{lithwick07}
{Lithwick}, Y., {Goldreich}, P., \& {Sridhar}, S. 2007, \apj, 655, 269

\bibitem[{{Maron} \& {Goldreich}(2001)}]{maron01}
{Maron}, J., \& {Goldreich}, P. 2001, \apj, 554, 1175

\bibitem[{{Matthaeus} \& {Goldstein}(1982)}]{matthaeus82a}
{Matthaeus}, W.~H., \& {Goldstein}, M.~L. 1982, \jgr, 87, 6011

\bibitem[{{Matthaeus} {et~al.}(1990){Matthaeus}, {Goldstein}, \&
  {Roberts}}]{matthaeus90}
{Matthaeus}, W.~H., {Goldstein}, M.~L., \& {Roberts}, D.~A. 1990, \jgr, 95,
  20673

\bibitem[{{Osman} \& {Horbury}(2007)}]{osman07}
{Osman}, K.~T., \& {Horbury}, T.~S. 2007, \apjl, 654, L103

\bibitem[{{Osman} \& {Horbury}(2009)}]{osman09a}
---. 2009, \ang, 27, 3019

\bibitem[{{Podesta}(2009)}]{podesta09a}
{Podesta}, J.~J. 2009, \apj, 698, 986

\bibitem[{{Podesta} \& {Bhattacharjee}(2009)}]{podesta09b}
{Podesta}, J.~J., \& {Bhattacharjee}, A. 2009, Phys. Rev. Lett., submitted
  (arXiv:0903.5041v1)

\bibitem[{{Robinson} \& {Rusbridge}(1971)}]{robinson71}
{Robinson}, D.~C., \& {Rusbridge}, M.~G. 1971, \pof, 14, 2499

\bibitem[{{Sahraoui} {et~al.}(2009){Sahraoui}, {Goldstein}, {Robert}, \&
  {Khotyaintsev}}]{sahraoui09}
{Sahraoui}, F., {Goldstein}, M.~L., {Robert}, P., \& {Khotyaintsev}, Y.~V.
  2009, \prl, 102, 231102

\bibitem[{{Schekochihin} {et~al.}(2009){Schekochihin}, {Cowley}, {Dorland},
  {Hammett}, {Howes}, {Quataert}, \& {Tatsuno}}]{schekochihin09}
{Schekochihin}, A.~A., {Cowley}, S.~C., {Dorland}, W., {Hammett}, G.~W.,
  {Howes}, G.~G., {Quataert}, E., \& {Tatsuno}, T. 2009, \apjs, 182, 310

\bibitem[{{Shebalin} {et~al.}(1983){Shebalin}, {Matthaeus}, \&
  {Montgomery}}]{shebalin83}
{Shebalin}, J.~V., {Matthaeus}, W.~H., \& {Montgomery}, D. 1983, \jopp, 29, 525

\bibitem[{{Taylor}(1938)}]{taylor38}
{Taylor}, G.~I. 1938, \rslpsa, 164, 476

\bibitem[{{Weygand} {et~al.}(2009){Weygand}, {Matthaeus}, {Dasso}, {Kivelson},
  {Kistler}, \& {Mouikis}}]{weygand09}
{Weygand}, J.~M., {Matthaeus}, W.~H., {Dasso}, S., {Kivelson}, M.~G.,
  {Kistler}, L.~M., \& {Mouikis}, C. 2009, \jgr, 114, 7213

\bibitem[{{Zweben} {et~al.}(1979){Zweben}, {Menyuk}, \& {Taylor}}]{zweben79}
{Zweben}, S.~J., {Menyuk}, C.~R., \& {Taylor}, R.~J. 1979, \prl, 42, 1270

\end{thebibliography}
\end{document}